\documentclass[conference, final]{IEEEtran} 
\usepackage{amsmath, amsthm, amssymb} 
\usepackage{epsfig}
\usepackage{color}
\usepackage[normalem]{ulem}
\usepackage{latexsym}
\usepackage{graphicx}
\usepackage{cite}

\usepackage{subfigure}

\title{Analytic Performance Evaluation of M${-}$QAM Based Decode-and-Forward Relay Networks Over Enriched Multipath Fading Channels}
\author{\IEEEauthorblockN{Mulugeta~K.~Fikadu}
\IEEEauthorblockA{Department of Electronics and \\ Communications Engineering\\
Tampere University   of \\ Technology (TUT) \\ FI-33101, Tampere, Finland \\ 
email:mulugeta.fikadu@tut.fi} 
\and
\IEEEauthorblockN{Paschalis~C.~Sofotasios}
\IEEEauthorblockA{Department of Electronics and \\ Communications Engineering\\
Tampere University  of \\ Technology (TUT) \\ FI-33101, Tampere, Finland  \\ 
email:paschalis.sofotasios@tut.fi} 
\and
\IEEEauthorblockN{Mikko Valkama}
\IEEEauthorblockA{Department of Electronics and \\ Communications Engineering\\
Tampere University  of \\ Technology  (TUT) \\ FI-33101, Tampere, Finland  \\ 
email:mikko.e.valkama@tut.fi} 
\and
\IEEEauthorblockN{Qimei Cui}
\IEEEauthorblockA{Wireless Technology \\ Innovation Institute\\
Beijing University of Posts \\ and Telecommunications \\ 100876, Beijing,  China \\ 
email:cuiqimei@bupt.edu.cn} }
\begin{document}

\maketitle

\begin{abstract}
The present work is devoted to the analysis of a regenerative multi-node dual-hop cooperative system over enriched multipath fading channels. Novel analytic expressions are derived for the  symbol-error-rate for $M{-}$ary quadrature modulated signals in  decode-and-forward relay systems over both independent and identically distributed as well as independent and non-identically distributed Nakagami${-}$q (Hoyt) fading channels. The derived expressions are based on the  moment-generating-function approach and  are given in closed-form in terms of the generalized Lauricella series.  The offered results are validated extensively through comparisons with respective results from computer simulations and are useful in the analytic performance evaluation of regenerative cooperative relay communication systems. To this end, it is shown that the performance of the cooperative system is, as expected, affected by the number of employed relays as well as by the value of the fading parameter $q$, which accounts for \textit{pre-Rayleigh} fading conditions that are often encountered in mobile cellular radio systems. 
\end{abstract}

\section{Introduction}

\pubidadjcol

Spatial diversity is among the most classical and effective methods  to overcome multipath fading in wireless channels. Multi-antenna communication systems constitute one method that realises spatial diversity; however, due to the required space and power constraints for efficient operation, this method is often considered relatively inefficient. Another, efficient technique to overcome such kind of resource limitations is cooperative or relay communications  \cite{SD, R14}.

Cooperative transmission  schemes exploit the broadcast nature and the inherent spatial diversity of fading channels without the need to install multiple antennas in a single mobile radio terminal, as in  conventional MIMO systems. In cooperative systems, wireless terminals share and coordinate their resources for relaying messages to each other  and for transmitting information signals over the numerous independent paths in the wireless network. Based on this, receivers exploit received signals often through different combining methods which have been extensively shown to provide efficient and robust operation in signal distortion by  fading effects. Cooperative communication systems can be typically realised by means of either  \textit{regenerative} or \textit{non-regenerative } methods.  The former typically refer to decode-and-forward (DF) relaying protocols whereas the latter refer to amplify-and-forward relaying protocols (AF), \cite{R18, New_1, New_2, New_3, New_4, R16} and the references therein. It has been shown that the digital processing nature of DF relaying makes it more practical than AF relaying which requires expensive RF transceivers to scale up  the analog signal for avoiding to additionally relay the noisy version of the signal \cite{SD},\cite{R14}.

It is widely known that fading phenomena have a significant effect on the performance of conventional and cooperative communications \cite{B:Alouini, New_5, New_6, New_7}. As a result, the limits of various communication scenarios have been studied by several researchers over the most basic multipath fading models. In this context, the authors in \cite{R18} proposed performance bounds of a DF multi-relay node networks over non-identical Nakagami${-}m$ fading channels. Specifically, the authors derived   upper and lower bounds for the outage probability (OP) for the case of maximum-ratio-combining (MRC) diversity. Likewise, two alternative lower tight bounds that approximate the corresponding OP were reported  in  \cite{R16} whereas the symbol error-rate (SER) of multi-node DF systems  over fading channels was addressed in \cite{R3, Trung, Paschalis_1, Paschalis_2}, and the references therein.

In spite of the numerous investigations on relay communications over fading channels, the majority of the analyses assume that multipath fading effects follow either the Rayleigh  or the Nakagami${-}m$ distribution. Nevertheless, the last years witnessed numerous advances in wireless channel characterization and modeling and in this context, the Nakagami${-}q$, or Hoyt, distribution  has been shown to provide accurate fading characterization of realistic indoor and outdoor multipath fading scenarios, particularly in the context of  mobile cellular radio systems \cite{Annamalai, Patzold}. The distinct feature of the Hoyt fading model is its capability to account for enriched multipath fading conditions which are characterized by  the absence of dominant components \cite{B:Alouini, Yacoub_1, Zogas, J:Paris_1, J:Tavares, J:Paris_2, J:Beaulieu, Yacoub_2}  and the references therein. Nevertheless, in spite of its proved usefulness, this model has not been widely investigated in the context of relay communications. Motivated by this, the present  work is devoted to the analytic performance evaluation of decode-and-forward systems over Hoyt distributed enriched multipath fading channels. Specifically, novel analytic  expressions are  derived for the case $M{-}$ary modulated signals over independent and identically distributed (i.i.d) and independent and non-identically distributed (i.n.i.d) Hoyt fading channels. The offered results are expressed in closed-form and their validity  is justified through comparisons with results from respective  computer simulations. As expected, it is shown that the performance of the considered system is highly dependant on both the number of employed relays and the value of the fading parameter $q$, which accounts for enriched fading conditions that are in the range of \textit{pre-Rayleigh} fading severity.  
   
The reminder of this paper is organized as follows: Section II presents the considered DF system and  channel model. The exact SER analysis for $M-$QAM modulated signals over i.i.d and i.n.i.d Hoyt channels is provided in Section III. The  corresponding performance analysis is provided in Section IV while closing remarks are given in Section V.  
\begin{figure}[!t]
\centering{\includegraphics[keepaspectratio,width= 2.5in]{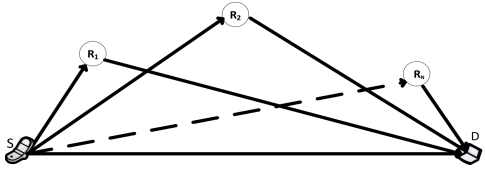}}
\caption{Multi-node dual hop cooperative relay network} 
\end{figure}
\section{Signal and System Models}

\subsection{Regenerative Relaying}

We consider a dual-hop cooperative radio access architecture with intermediate nodes  between a source   $S$ and a destination   $D$  as illustrated in Fig. $1$. Each relay node is equipped with a single antenna assuming  half-duplex communication in the context of a decode and forward cooperative protocol. Furthermore, the nodes in the system transmit signals through orthogonal channels for avoiding inter-relay interference using for example time division multiple access (TDMA). Based on this, in phase I the source broadcasts the information signal to the destination and to the relay nodes which is  expressed as,  
 \begin{equation}\label{L1}
 y_{S,D} = \sqrt{P_{S}} h_{S,D} x + n_{S,D}
  \end{equation}
  and 
    \begin{equation}\label{L2}
  y_{S,R_k} = \sqrt{P_{S}} h_{S,R_k} x + n_{S,R_k}, \quad k  \in \left\lbrace {1,2,3, . . . , K}\right\rbrace 
  \end{equation}
respectively,  where $P_{S}$ denotes the transmit source power, $x$ is the transmitted symbol with normalized unit energy, $ h_{S,D}$ and $ h_{S,R_k}$ are the complex channel gains of the source-destination and source-relay  links, and $ n_{S,D}$ and $n_{S,R_k}$ are the corresponding additive-white Gaussian noise (AWGN) terms with zero mean and variance $ N_{0} $. In the next $k +1$ available time slot, if a relay $ k $ decodes correctly,  it forwards the information signal to the destination with power $  \bar{P}_{R} = P_{R} $. Otherwise, if the decoding is unsuccessful the relay remains silent i.e., $\bar{P}_{R} = 0 $. Based on this, the received signal at the destination terminal is given by,  
\begin{equation}\label{L3}
y_{R_k,D} = \sqrt{\bar{P}_{R_k}} h_{R_k,D} x + n_{R_k,D},  \quad k \in  \left\lbrace { 1, 2, 3, . . . , K }\right\rbrace  
\end{equation}
where $h_{R_{k,D}}$ is the complex channel gain from the $k^{\rm th}$ relay to-destination and $n_{R_{k,D}}$ is the corresponding  AWGN. Using the maximal-ratio-combining technique, the destination node coherently combines the signals received from the $K{-}$relays and the source as follows \cite{YL}:  
\begin{equation}\label{L4}
y_{D} = w_{1}y_{S,D} + \sum \limits_{k =1}^K w_{2}y_{R_k,D} 
\end{equation}
where $w_{1} = \sqrt{P_{s}}h^{*}_{S,D}/ N_{0}$  and  $ w_{2} = \sqrt{\bar{P}_{R_k}}h^{*}_{R_k,D}/N_{0}$ are the MRC coefficients for the  $y_{S,D}$ and $y_{R_k,D}$ signals, respectively.

\subsection{Nakagami${-}q$ (Hoyt) Multipath Fading}

As already mentioned, the Hoyt  fading model has been shown to represent effectively enriched  in non-line-of-sight (NLOS) communication scenarios.  The  PDF of the instantaneous SNR is given by  \cite[eq. (2.11)]{B:Alouini}, namely, 
\begin{equation} \label{PDF_Hoyt} 
p_{\gamma}(\gamma) = \frac{1 + q^{2}}{2q \overline{\gamma}} {\rm exp} \left(- \frac{(1 + q^{2})^{2} \gamma}{4 q^{2}\overline{\gamma}}\right) I_{0}\left( \frac{(1 - q^{4}) \gamma}{4 q^{2} \overline{\gamma}} \right)  
\end{equation}
where $ \overline{\gamma} = E(\gamma) = E ( \; \left| h \right|^{2})P/N_{0} = \Omega P/N_{0} $ represents the average SNR with  $ E (\cdot) $ denoting statistical expectation.    The corresponding CDF and MGF expressions are given by \cite[eq. (9)]{J:Paris_1} and \cite[eq. (2.12)]{B:Alouini}, respectively, namely,
\begin{equation} \label{CDF}
\begin{split}
P_{\gamma}(\gamma) &= Q_{1}\left( \sqrt{\frac{(1 + q^{4}) (1 + q) \gamma}{8q (1-q) \overline{\gamma} }}, \sqrt{\frac{(1 - q^{4}) (1 - q) \gamma}{8q (1+q) \overline{\gamma} }}\right)  \\
&- Q_{1}\left( \sqrt{ \frac{(1 - q^{4}) (1 - q) \gamma}{8q (1+q) \overline{\gamma} }}, \sqrt{\frac{(1 + q^{4}) (1 + q) \gamma}{8q (1-q) \overline{\gamma} }}\right)  
\end{split}
\end{equation} 
and
\begin{equation}\label{L5}
M_{\gamma}(s) = \left(1-2s\overline{\gamma} + \frac  {(2s\overline{\gamma})^{2}q^{2}}{(1 + q^{2})^2}     \right)^{-\frac{1}{2}}
\end{equation}
where $Q_{1}(a,b)$ denotes the Marcum $Q{-}$function \cite{pud, New_8, New_9}. Notably, the above expressions have a convenient and easily computed  mathematical representation.  
It is also recalled   that  the fading severity of  Hoyt fading channels is inverse proportional to the value of the fading parameter $q$ and it includes Rayleigh distribution as a specific case for $q = 1$, \cite{B:Alouini}.

\section{ Exact Symbol Error Rate Analysis }

The average symbol-error-rate for the considered K-relays dual-hop cooperative  system  is given by \cite{R3, YL}, 
\begin{equation}\label{L6}
P_{SER}^{D}  = \sum\limits_{z = 0}^{2^{K}-1} P(e|\mathbf{B} = E_{z})P(\mathbf{B}= E_{z})
\end{equation}
where the vector $ \mathbf{B} = [B(1),B(2), B(3), \ldots , B(K)]$ accounts for the state of relay nodes with   $ B(k)$  representing the state of a each relay and takes values $1$ and $0$ for the case of successful and unsuccessful decoding, respectively.  Furthermore,  $ \mathbf{E_{z}} = [E(1),E(2), E(3), \ldots , E(K)]$ denotes different possible combinations of decoding results of the relay nodes where $z$ is a particular decoding result combination ranging from $ 0 \; \text{to} \; 2^{K}-1$. Also, $ P(e|\mathbf{B} = E_{z}) $ denotes the conditional error probability whereas $ P(\mathbf{B} = E_{z}) $  is the corresponding probability of the relay node decoding outcomes. 
Thus, for the case of statistically independent channels, the joint probability of particular combination of state values can be expressed as $ \mathbf{P(B)} = P(B(1)) P(B(2)) P(B(3))  \ldots P(B(K)).$  

In the case of MRC in \eqref{L4}, the instantaneous SNR for a given particular $E_{z}$ at the destination is expressed as \cite{YL}
\begin{equation}\label{7}
\gamma_{MRC} (E_{z})   =  \frac{P_{S}\, |h_{S,D}|^{2}  +  \sum \limits _{k = 1}^K E_{z}P_{R_k}|h_{R_k,D}|^{2}}{N_{0}}
\end{equation}
with the corresponding  MGF given by  \cite{SA},
\begin{equation}\label{L8}
M_{\gamma_{MRC}}(s) = M_{\gamma_{S,D}}(s)\prod\limits_{k=1}^{K}E_{z}M_{\gamma_{R_{k},D}}(s). 
\end{equation} 
Importantly, the MGF in \eqref{L5} can be  equivalently expressed in terms of  the Craig representation, namely,  
\begin{equation}\label{L9}
 M_{\gamma}\left(-\frac{g}{\sin^{2}\theta }\right) =  \frac{1}{\sqrt{\left(1+\frac{2\bar{\gamma}g}{(1+q^2)\sin^{2}\theta}\right)\left(1+\frac{2\bar{\gamma} q^2 g}{(1+q^2)\sin^{2}\theta}\right)}}.
\end{equation}
To this effect, for the case $M-$QAM mapping one obtains, 
\begin{equation}\label{L11}
\begin{split}
P(\gamma_{MRC}) &= \frac{4C}{\pi}\underbrace{\int_{0}^ {\pi/2}M_{\gamma_{MRC}}\left(-\frac{g_{QAM}}{\sin^{2}\theta}\right)d\theta}_{\triangleq \: \mathcal{I}_{1}} \\ 
& -\frac{4C^{2}}{\pi}\underbrace{\int_{0}^ {\pi/4}M_{\gamma_{MRC}}\left(-\frac{g_{QAM}}{\sin^{2}\theta}\right)d\theta }_{\triangleq \:\mathcal{I}_{2}}
\end{split}
\end{equation}
where $C = (1 - 1/\sqrt{M})$ and  $ g_{QAM} = 3/2(M-1)$. Evidently,  determining the above decoding error probability is subject to analytic evaluation of the $ \mathcal{I}_{1} \;\text{and}\; \mathcal{I}_{2}$ integrals  in \eqref{L11}.  

\subsection{The case of i.n.i.d   Hoyt fading channels}
  
This subsection is devoted to the derivation of a  the closed-form expression for the error probability for the generic case that the value of $q$ and $\gamma$ is not necessarily equal in each wireless link. To this end, with the aid of \eqref{L8} and \eqref{L9} the $ \mathcal{I}_{1} $ term can be expressed   as follows: 
\begin{equation}\label{L13}
\hspace{- 0.2cm}  {\mathcal{I}_{1}}   = \int\limits_{0}^ {\pi/2} \prod_{k = 1}^{K}   \frac{E_{z} \,  d\theta}{\sqrt{\left(1+\frac{A_{1}}{\sin^{2}\theta}\right)\left(1+\frac{A_{2}}{\sin^{2}\theta}\right)\left(1+\frac{B_{1k}}{\sin^{2}\theta}\right)\left(1+\frac{{B_{2k}} }{\sin^{2}\theta}\right)}}     
\end{equation}
where 
\begin{equation}
{A_{1} = \frac{2g_{QAM}\overline{\gamma}_{S,D}}{ 1 + q^{2}_{S,D} }},
\end{equation}
\begin{equation}
 A_{2}= \frac{2q^{2}_{S,D}g_{QAM} \overline{\gamma}_{S,D}}{  1 + q^{2}_{S,D} },
\end{equation}
\begin{equation}
{B_{1k} = \frac{2g_{QAM}\overline{\gamma}_{R_k,D}}{  1 + q^{2}_{R_k,D} }},
\end{equation}
and
\begin{equation}
B_{2k} = \frac{2g_{QAM} q^{2}_{R_k,D}\overline{\gamma}_{R_k,D}}{   1 + q^{2}_{R_k,D} }. 
\end{equation}
 By letting $ u = \sin^{2}(\theta) $ in \eqref{L13} and therefore,  $d{/}du = 2\cos(\theta) \sin(\theta)$ and    $\cos(\theta) = \sqrt{1 - u}$, it  follows that,  
\begin{equation}\label{L14}
\hspace{- 0.1cm}  \mathcal{I}_{1}   =  \int_{0}^ {1}  \frac{u^{E_{z}K \, + \frac{1}{2}  } (1 - u)^{- \frac{1}{2} }}{2\left( u + A_{1}\right)^{\frac{1}{2}}\left( u + A_{2}\right )^{\frac{1}{2}}} \prod_{k = 1}^{K} \frac{E_{z}  \, du}{\left( u +B_{1k} \right)^{\frac{1}{2}}\left( u+ B_{2k}\right  )^{\frac{1}{2}}}.    
\end{equation} 
By factoring out the constant terms, the above integral can be explicitly written as in \eqref{LL15} at the top of the next page.  
\begin{figure*}
\begin{equation}\label{LL15}
{\mathcal{I}_{1}}  =   \frac{1}{ 2 \sqrt{A_{1}A_{2}} \prod_{k=1}^{K} \sqrt{E_{z}B_{1k}B_{2k}}}\int_{0}^ {1}  \frac{u^{1/2 + E_{z}K} (1 - u)^{-1/2}}{ \left(1 + \frac{u}{A_{1}}\right)^{1/2}   \left(1 + \frac{u}{A_{2}}\right)^{1/2}}    \prod_{k = 1}^{K}  \frac{1}{ \left(1 + \frac{u}{B_{1k}}\right)^{1/2}  \left(1 + \frac{u}{B_{2k}}\right)^{1/2}}du. 
\end{equation} 
\hrulefill
\end{figure*}
It is recalled here that the channel fades and powers in each wireless link are non-identical i.e.   $ q_{R_1,D} \neq  q_{R_2,D} \neq  \cdots  \neq  q_{R_K,D}$ and $ \overline{\gamma}_{R_1,D} \neq \overline{\gamma}_{R_2,D} \neq \cdots \neq  \overline{\gamma}_{R_K,D} $. Based on this and after basic algebraic manipulations, a closed-form expression is  deduced  for $\mathcal{I}_{1}$ in  \eqref{LLL18}, at the top of the next page. 
\begin{figure*}
\begin{equation}\label{LLL18}
{\mathcal{I}_{1}}  =   \frac{  \sqrt{\pi} \Gamma \left( \frac{3}{2} + E_{z}K \right)F^{(2E_{z}K + 2)}_{D} \left( \frac{3}{2} + E_{z}K; \overbrace{\frac{1}{2}, \frac{1}{2}, \cdots , \frac{1}{2}}
  ^{2K + 2}; 2 + E_{z}K; - \frac{1}{A_1}, - \frac{1}{A_2}, \overbrace{- \frac{1}{B_{11}},\cdots ,- \frac{1}{B_{1K}},-\frac{1}{B_{21}},\cdots,-\frac{1}{B_{2K}}}^{2K}  \right)  }{ 2 \sqrt{A_{1}A_{2}} \Gamma \left( 2 + E_{z}K \right) \prod_{k = 1}^{K}\sqrt{E_{z}B_{1k}B_{2k}} }. 
\end{equation} 
\hrulefill
\end{figure*}
where, 
\begin{equation}\label{LL17}
\begin{split}
F_{D}^{(n)}(&a;b_{1},b_{2},b_{3},\ldots , b_{n};c;x_{1},x_{2},x_{3},\ldots,x_{n}) \triangleq \\ 
& \frac{\Gamma{(c)}}{\Gamma{(a)}\Gamma{(c-a)}}\int_{0}^{1} \frac{t^{a-1}(1-t)^{c-a-1}}{(1-x_{1}t)^{b_{1}} \ldots (1-x_{n}t)^{b_{n}}} dt 
\end{split}
\end{equation} 
is the generalized Lauricella function   of $n$ variables   \cite{pud, NYY}. 
Importantly, the integrals $\mathcal{I}_{1}$ and $\mathcal{I}_{2}$ have the same integrand and differ in the upper limit. Therefore, by following exactly the same methodology and additionally letting $y = 2u$ leads to \eqref{LLL19}, which yields straightforwardly   the closed-form expression for $\mathcal{I}_{2}$ in  \eqref{LL20} at the top of the next page.  
\begin{figure*}
\begin{equation}\label{LLL19}
\mathcal{I}_{2}  =  \frac{1}{2^{(EzK + 5/2)} \sqrt{A_{1}A_{2}}  \prod_{k=1}^{K} \sqrt{E_{z}B_{1k}B_{2k}} }  \int_{0}^ {1}  \frac{y^{1/2 + E_{z}K}(1-\frac{y}{2})^{-1/2}}{\left(1+\frac{y}{2A_{1}}\right)^{1/2}\left (1+\frac{y}{2A_{2}}\right)^{1/2}}    \prod_{k = 1}^{K} \frac{1}{\left(1+\frac{y}{2B_{1k}}\right)^{1/2}\left(1+\frac{y}{2B_{2k}}\right)^{1/2}}dy.  
\hrulefill
\end{equation}
\end{figure*}
\begin{figure*}
\begin{equation}\label{LL20}
\hspace{- 0.25cm} \mathcal{I}_{2}  = \frac{\Gamma(\frac{3}{2}+E_{z}K)\;F_{D}^{(2E_{z}K +3)}\left(\frac{3}{2}+E_{z}K; \overbrace{\frac{1}{2}, \frac{1}{2}, \cdots, \frac{1}{2}}^{2K+3}; \frac{5}{2}+E_{z}K; - \frac{1}{2A_{1}}, - \frac{1}{2A_{2}}, \overbrace{- \frac{1}{2B_{11}},\cdots, - \frac{1}{2B_{1K}}, - \frac{1}{2B_{21}},\cdots, - \frac{1}{2B_{2K}}}^{2K},  \frac{1}{2}   \right)}{2^{(EzK + 5/2)}  \sqrt{A_{1}A_{2}}\;\Gamma(\frac{5}{2}+E_{z}K)\prod_{k = 1}^{K}{\sqrt{E_{z}B_{1k}B_{2k}}}}.
\end{equation}
\hrulefill
\end{figure*}

\subsection{The case of i.i.d   Hoyt fading channels}

The case of identical links  assumes   $ q_{R_1,D} =  q_{R_2,D} =  \cdots  =   q_{R_K,D} = q \; \text{and}\;  \overline{\gamma}_{R_1,D} =  \overline{\gamma}_{R_2,D} =  \cdots =  \overline{\gamma}_{R_K,D} = \overline{\gamma} $, and therefore,   $  B_{11} =  B_{12} = \cdots =  B_{1K} =  B_{1} \;\text{and}                \; B_{21} = B_{22} \cdots =  B_{2K} = B_{2}$. Based on this and having derived the general case of non-identically distributed links, the $\mathcal{I}_{1}$ and $\mathcal{I}_{2}$ for the i.i.d case can be straightforwardly deduced as a special case yielding \eqref{LL21} and \eqref{LL22}, respectively,  at the top   page five.   
\begin{figure*}
\begin{equation}\label{LL21}
\mathcal{I}_{1}  =   \frac{  \sqrt{\pi} \Gamma \left( \frac{3}{2} + E_{z}K \right)F^{(4)}_{D} \left( \frac{3}{2} + E_{z}K; \frac{1}{2}, \frac{1}{2}, \frac{E_{z}K}{2},\frac{E_{z}K}{2}
  ; 2 + E_{z}K; - \frac{1}{A_1}, -\frac{1}{A_2}, - \frac{1}{B_{1}}, -\frac{1}{B_{2}}  \right)  }{ 2 \sqrt{A_{1}A_{2}} \Gamma \left( 2 + E_{z}K \right) (B_{1}B_{2})^{E_{z}K/2} }. 
\end{equation}
\hrulefill
\end{figure*}
\begin{figure*}
\begin{equation} \label{LL22}
\mathcal{I}_{2}  = \frac{\Gamma(\frac{3}{2}+E_{z}K)\;F_{D}^{(5)}\left(\frac{3}{2}+E_{z}K; \frac{1}{2},\frac{1}{2}, \frac{E_{z}K}{2},\frac{E_{z}K}{2}  ,\frac{1}{2}; \frac{5}{2}+E_{z}K; - \frac{1}{2A_{1}}, - \frac{1}{2A_{2}} - \frac{1}{2B_{1}},- \frac{1}{2B_{2}},  \frac{1}{2}   \right)}{2^{(EzK + 5/2)}  \sqrt{A_{1}A_{2}}\;\Gamma(\frac{5}{2}+E_{z}K)(B_{1}B_{2})^{E_{z}K/2}}.
\end{equation}
\hrulefill
\end{figure*}

It is evident that with the aid of the  derived closed-form expressions for $\mathcal{I}_{1}$ and $\mathcal{I}_{2}$, for both i.n.i.d and i.i.d scenarios, the conditional error probability at the destination terminal after the MRC can be straightforwardly determined by, 
\begin{equation}\label{LL23}
P(e|\mathbf{B} = E_{z}) = \frac{4C}{\pi}\mathcal{I}_{1} - \frac{4C^{2}}{\pi}\mathcal{I}_{2}.  
\end{equation}
In order to derive a closed-form expression for the overall SER of the considered system, we additionally need to determine the decoding probability of the relay nodes $ P(\mathbf{B} = E_{z})$ which is a direct product of the element terms $ P (\overline{\gamma}_{S,R_k})$ i.e. decoding error at the relay nodes $R_{k}$ and $ (1- P (\overline{\gamma}_{S,R_k}))$. This is  also obtained by applying the aforementioned MGF approach for the source to relay nodes links, namely,  
\begin{equation} \label{LLL24}
\begin{split}
P(\gamma_{S,R_k}) &= \frac{4C}{\pi}\underbrace{\int_{0}^ {\pi/2}M_{\gamma_{S,R_k}}\left(-\frac{g_{QAM}}{\sin^{2}\theta}\right)d\theta}_{\triangleq \; \mathcal{I}_{3}} \\ 
&-\frac{4C^{2}} {\pi}\underbrace{\int_{0}^ {\pi/4}M_{\gamma_{S,R_k}}\left(-\frac{g_{QAM}}{\sin^{2}\theta}\right)d\theta}_{\triangleq \;\mathcal{I}_{4} }.
\end{split}
\end{equation} 
In order to evaluate $\mathcal{I}_{3}$ and $\mathcal{I}_{4}$ in  closed-form, we follow the same procedure as in the derivation of the closed-form solutions  for $\mathcal{I}_{1}$ and $\mathcal{I}_{2}$. This is achieved thanks to the similar algebraic formulation and representation, except from the fact that there is no involvement of $ q_{R_k,D}$. Based on this and after long but basic algebraic manipulations, it immediately follows that the $\mathcal{I}_{3}$ and $\mathcal{I}_{4}$ integrals  can be expressed as, 
\begin{equation}\label{LL25}
\mathcal{I}_{3} = \frac{\pi}{4\sqrt{C_{1}C_{2}}}F_{D}^{(2)}\left( \frac{3}{2};\frac{1}{2}, \frac{1}{2};2; -\frac{1}{C_{1}},    -\frac{1}{C_{2}}\right)
\end{equation}
and
\begin{equation}\label{LL26} 
\mathcal{I}_{4} = \frac{1}{6\sqrt{2C_{1}C_{2}}}F^{(3)}_{D}\left( \frac{3}{2};\frac{1}{2}, \frac{1}{2},\frac{1}{2};\frac{5}{2};- \frac{1}{2C_{1}}, - \frac{1}{2C_{2} },  \frac{1}{2}  \right)
\end{equation}
respectively, where 
\begin{equation}
C_{1}=\frac{2g_{QAM}\overline{\gamma}_{S,R_k}}{ 1 + q^{2}_{S,R_k} }
\end{equation}
and
\begin{equation}
C_{2}=\frac{2q^{2}_{S,R_k}g_{QAM}\overline{\gamma}_{S,R_k}}{  1 + q^{2}_{S,R_k} }.
\end{equation}
To this effect,    the decoding error probability of the relay nodes can be readily obtained as follows: 
\begin{equation}\label{LL27}
P(\gamma_{S,R_k}) = P(B = 0) = \frac{4C}{\pi} \mathcal{I}_{3}  - \frac{4C^{2}}{\pi} \mathcal{I}_{4}.
\end{equation}

It is recalled here that the  SER of the considered system is given in \eqref{L6} and is obtained with the aid of the derived results on the  probability for decoding error at the destination after the MRC, which is computed in terms of $ \mathcal{I}_{1}$ and $ \mathcal{I}_{2}$, and the individual decoding probabilities of the nodes for a given combination $ \mathbf{E_{z}}$. As a result, a novel analytical solution is obtained for the average SER of the cooperative network over   Nakagami${-}q$ (Hoyt) fading channels. It is also noted that the offered analytic expressions are given in terms of well known functions that are defined by single finite integrals which consist of elementary functions and can be computed using  scientific software packages such as MAPLE and MATLAB.

\section{Numerical Results}

This section is devoted to the  analysis of the offered results on the SER of the considered model for different communication scenarios.  
The variance of the noise is normalized to unity i.e. $N_{0} = 1 $ while  square M${-}$QAM constellation is employed assuming equally allocated  transmit powers to the source and the relay nodes. Furthermore, the average   power of the wireless links ${\Omega_{i,j}}$ is   assumed equal to    unity.  

\begin{figure}[t!]
\centering{\includegraphics[keepaspectratio, width= 3.35in]{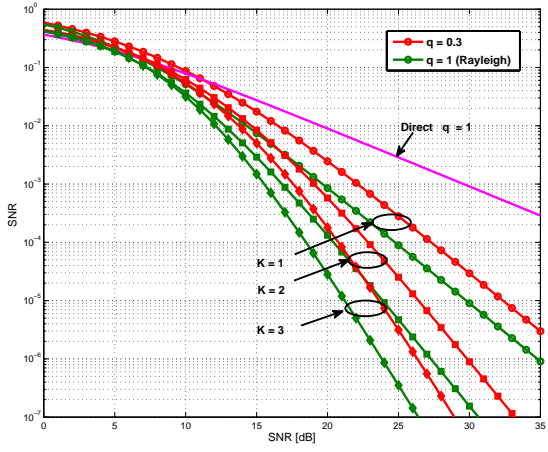} }
\caption{SER performances over Nakagami${-}q$ (\textit{Hoyt}) fading channels for $ q = 0.3$, $q = 1$ with  $\Omega_{S,D} = \Omega_{S,R_k} = \Omega_{R_k,D} = 1 $  \, for 4${-}$QAM Signals with different number of relays and direct reference.} 
\end{figure}

\begin{figure}[!t]
\centering{\includegraphics[keepaspectratio, width= 3.35in]{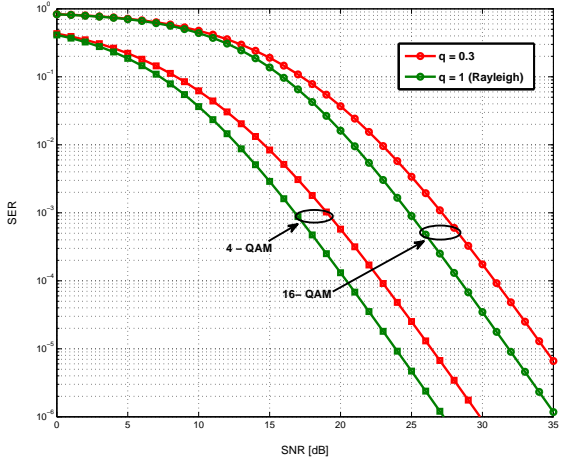} }
\caption{SER performances over Nakagami${-}q$ \;(\textit{Hoyt})\; fading channels $ q = 0.3$, $q = 1$ and \; $\Omega_{S,D} = \Omega_{S,R_k} = \Omega_{R_k,D} = 1 $  for 4${-}$QAM  and 16${-}$QAM Signals for the case of two relays.} 
\end{figure}

\begin{figure}[!t]
\centering{\includegraphics[keepaspectratio, width= 3.75in]{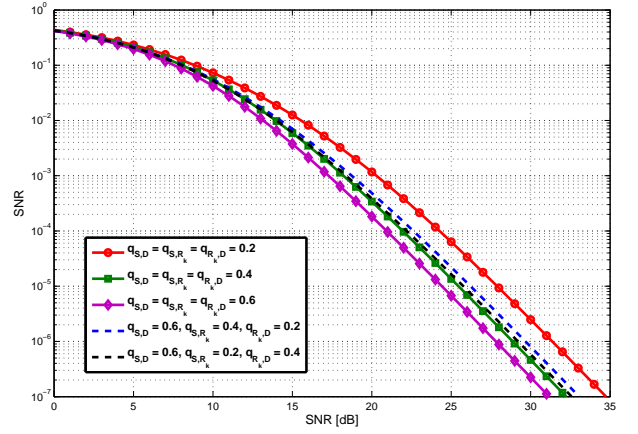} }
\caption{SER performances over Nakagami${-}q$ \;(\textit{Hoyt})\; fading channels with different fading parameters $ q_{S,D},q_{S,R_k},q_{R_k,D}$ and  \; $\Omega_{S,D} = \Omega_{S,R_k} = \Omega_{R_k,D} = 1 $\; for  4${-}$QAM Signals employing two relays.} 
\end{figure}

We firstly plot the corresponding SER as a function of SNR employing one, two and three relay nodes for symmetrical channel scenarios. In Fig. 2, the Hoyt fading parameter is indicatively set to $ q = 0. 3$ and $q = 1.0 $ while all average channel fading powers (channel variances) are   assumed to be unity. One can observe the effect of the variation of the severity of enriched fading on the overall system performance. This figure also includes the performance for the case of   direct transmission for $q = 1$ (Rayleigh fading) as a reference for demonstrating  the effect of enriched fading.   In this context, it is also shown that, as expected,  the dual hop regenerative communication systems improves significantly the overall system performance regardless of the  limited number of relays, considered in this case.

In the same context, Fig. 3 illustrates the SER for different modulation order, namely, $M = 4$ and $M = 16$ in the case of both severe and Rayleigh  fading conditions i.e. $q = 0. 3$ and $q = 1. 0$, respectively. One can notice that the respective SERs differ to each-other for one or even two orders of magnitude in the moderate and high SNR regime.  Likewise,  Fig. 4 demonstrates  the cooperation performance of $4{-}$QAM system  of two relay scenario over    \textit{Hoyt} fading channels for different fading  parameters of $ q_{S,D},q_{S,R_k}\;\text{and}\; q_{R_k,D} $. By varying the value of  $q_{i,j}$,we observe from the plots the effect of $q$ on the symbol  error rates over the radio fading channels. Based on this, we can verify that increasing the values of $q$ in the symmetric channel condition improves the performance, i.e. minimizes the SER, of the cooperation system. In particular, as the fading  parameter approached unity, the performance of the system approached gradually the respective Rayleigh performance. In addition, it is shown that the SER degradation is  affected more by  the value of  $ q_{R_k,D} $ than by the value of $q_{S,R_k}$ and thus, possessing knowledge of the fading characteristics of the wireless links is rather important in the design and deployment of regenerative relay systems.

\section{Conclusion}

This paper analyzed the performance of  multi-node dual-hop regenerative relay wireless system  in the presence of  enriched multipath fading conditions. Both the case of independent and identically distributed channels and independent and non-identically distributed channels were considered and novel closed-form expressions were derived for the corresponding symbol error rate for the case of $M{-}$QAM modulated signals employing maximum ratio combining. The involved analysis was based on the MGF approach and it was shown that the system performance is affected by enriched fading conditions, particularly for small values of the Hoyt fading parameter, regardless of the number of employed relay nodes.   In general, the provided analytical results can be used to analyze the  performance of general M${-}$QAM based decode-and-forward relay networks over Nakagami${-}q$ (Hoyt) fading channels.


{}

\end{document}